\documentclass[11pt]{article}

\usepackage{graphicx,amsmath,amsmath,amsfonts,amssymb,dcolumn,amsthm,euscript,soul,float}
\usepackage{hyperref}
\usepackage{xcolor}
\begin{document}

% \usepackage{setspace}
% \usepackage{color}
% \usepackage{slashed}
% \usepackage{comment}
% \usepackage{citesort}
%\setcounter{secnumdepth}{2} 

%\geometry{left=1.5cm, right=1.5cm, top=1.5cm, bottom=1.5cm}
%\setlength{\topmargin}{-1cm} \setlength{\evensidemargin}{-0.75cm}
%\setlength{\oddsidemargin}{-0.75cm} \setlength{\textwidth}{17.5cm}
%\setlength{\textheight}{22.5cm} \setlength{\parskip}{10pt}

%----------------------------------------------------------------------------------------
%	Titulo
%----------------------------------------------------------------------------------------

\date{}
\title{\textbf{Renormalizability of the center-vortex free sector of Yang-Mills theory
}}

%----------------------------------------------------------------------------------------
%	Autores e informa??es
%----------------------------------------------------------------------------------------

 \author{ \textbf{D.~Fiorentini}\thanks{diego\_fiorentini@id.uff.br}\,\,,
 \textbf{D.~R.~Junior}\thanks{davidjunior@id.uff.br}\,\,,
  \textbf{L.~E.~Oxman}\thanks{leoxman@id.uff.br}\,\,,
 \textbf{R.~F.~Sobreiro}\thanks{rodrigo$\_$sobreiro@id.uff.br}\,\,,\\
 {\small   \textnormal{  \it
 UFF -- Universidade Federal Fluminense,}}\\
{\small \textnormal{ \it Instituto de F\'{\i}sica, Campus da Praia Vermelha,}}\\
\small \textnormal{ \it Avenida Litor\^anea s/n, 24210-346, Niter\'oi, RJ, Brasil.}\\
} 	

\maketitle

%----------------------------------------------------------------------------------------
%	ABSTRACT
%----------------------------------------------------------------------------------------

\begin{abstract}
In this work, we analyzed a recent proposal to detect $SU(N)$ continuum Yang-Mills sectors labeled by center vortices, inspired by Laplacian-type center gauges in the lattice. Initially, after the introduction of appropriate external sources, we obtained a rich set of sector-dependent Ward identities, which can be used to control the form of the divergences. Next, we were able to show the all-order multiplicative renormalizability of the center-vortex free sector. These are important steps towards the establishment of a first principles, well-defined, and calculable Yang-Mills ensemble.

\end{abstract}

%----------------------------------------------------------------------------------------
\section{Introduction}
%---------------------------------------------------------------------------------------- 

As is well-known, the Fadeev-Popov procedure to quantize Yang-Mills (YM) theories \cite{Faddeev:1967fc}, so successful to make contact with experiments at high energies, cannot be extended to the infrared regime \cite{Gribov:1977wm,Sobreiro:2005ec}. In covariant gauges, this was established by Singer's theorem \cite{Singer:1978dk}: for any gauge-fixing, there are orbits with more than one gauge field satisfying the proposed condition. In a geometrical language,  the YM principal bundle has a topological obstruction which makes it impossible to find a global section $f$ in the complete gauge field  configuration space $\{ \mathcal{A}_\mu \}$.  
 
A way to cope with this problem was extensively studied in the last decades, mostly in the Landau and linear covariant gauges. This is the (refined) Gribov-Zwanziger (GZ) approach \cite{Zwanziger:1988jt,Zwanziger:1989mf,Zwanziger:1992qr,Dudal:2005na,Dudal:2008sp,Dudal:2011gd}, where the configurations to be path-integrated are restricted so as to avoid infinitesimal copies. In this respect, it is worth mentioning that this region is not free from copies, since it still contains those related with finite gauge transformations. In fact, the existence of finite Gribov copies inside the Gribov region was discussed in  \cite{VANBAAL1992259} and references therein, where it is argued about the relation between the Gribov region and the fundamental modular region, the latter being free of copies.  Along this line, a BRST (Becchi, Rouet, Stora and Tyutin) invariant action was obtained in Euclidean spacetime \cite{Capri:2016aqq,Capri:2018ijg}, which provides a calculational tool similar to the one used in the perturbative regime. Beyond the linear covariant gauges, many efforts were also devoted to  the maximal Abelian gauges, see Ref. \cite{Capri:2015pfa} and references therein. The BRST invariance is an important feature to have predictive power (renormalizability), as well as to show independence of observables on gauge-fixing parameters. Another interesting feature of the GZ approach is that the gluon Green's functions get drastically modified in the infrared,  pointing to a destabilization of the perturbative regime. The perturbative pole that would correspond to an asymptotic massless particle is replaced by an infrared suppressed behavior. However, as the Green's functions are not gauge invariant objects, it is not clear how to define gluon confinement. A related discussion is regarding the presence of complex poles in the gluon Green's functions, which led to look for correlators of gauge invariant operators only displaying physical poles \cite{Dudal:2010cd}. In addition to the GZ treatment, alternative analytical approaches have been used to determine correlation functions displaying complex poles, see for instance \cite{Tissier:2010ts,Tissier:2011ey,Pelaez:2014mxa,Siringo:2019lmg,Siringo:2018uho}.
 
A deeper issue is how to get closer to the confinement of quark probes in pure YM theories, whose order parameter is the Wilson loop. This phenomenon is not only associated to a linearly rising potential but to the formation of a flux tube (see \cite{10.1093/ptep/ptz093,cosmai} and references therein) with transverse collective modes 
 \cite{LUSCHER1981317,ATHENODOROU2007132}. This is a physical object on its own, which is well beyond the language of Green's functions and the usual Feynman diagrams. Then, a fundamental question is if there is a first principles formulation of YM theories which allows to envisage a connection with a confining flux tube.    
 
In this regard,  since the very beginning of gauge theories, we were used to pick a gauge fixing and then do the calculations for this choice. When confronted with Singer's theorem, this naturally led to a restriction on $\{ \mathcal{A}_\mu \}$ to eliminate the associated Gribov copies. However, we could try to resolve the obstruction possed by this theorem in a different way. While Singer's theorem states that it is impossible to construct a global section  $f$, it does not rule out the possibility of  covering $\{ \mathcal{A}_\mu \}$  with local regions $\mathcal{V}_\alpha$, each one having its own well-defined local section $f_\alpha$,  for $\mathcal{A}_\mu \in \mathcal{V}_\alpha $.
 
In Ref. \cite{Oxman:2015ira}, an approach motivated by lattice center gauges was proposed in the continuum. There, a tuple of auxiliary adjoint fields $\psi_I \in \mathfrak{su}(N)$, with $I$ being a flavor index, were initially introduced by means of an identity,  thus keeping the pure YM dynamics unchanged.   
The identity  was constructed using equations of motion that correlate $\mathcal{A}_\mu$ with $\psi_I$. Next, a polar decomposition in terms of a ``modulus'' tuple $q_I$ and phase $S$ was applied to the fields $\psi_I$ 
\begin{equation}
\psi_I   = S q_I S^{-1}    \makebox[.5in]{,}   S \in SU(N) 
\label{m-p}
\end{equation}
(for the definition of modulus, see Sec. \ref{sec-q}). 
As this was done  covariantly, a gauge transformed field $\mathcal{A}_\mu^U$ is mapped to
\begin{equation}
S[\mathcal{A}^U] = U S[\mathcal{A}] \;.
\label{covar}   
\end{equation}
 Although $\mathcal{A}_\mu$ is smooth, $S[\mathcal{A}]$ generally contains defects, which cannot be eliminated by gauge transformations. This makes it impossible to define a global reference phase. Instead, we have to split $\{ \mathcal{A}_\mu \}$  into sectors $\mathcal{V}(S_0)$,  formed by those $\mathcal{A}_\mu$ that can be gauge transformed to some $A_\mu$ with $S[A]=S_0$.  The condition for copies in the orbit of $A_\mu$ is $ S[A^U] = S_0$, which due to Eq. \eqref{covar} implies $ U \equiv  I$. Therefore, on each sector, there are no copies. Indeed, the sectors $\mathcal{V}(S_0)$ provide a partition of the configuration space:
\begin{align}
\{ \mathcal{A}_\mu \} =  \mathbin{\scalebox{1.5}{\ensuremath{\cup}}}_{S_0}  \,  \mathcal{V}(S_0)\;, \makebox[.5in]{} \mathcal{V}(S_0) \mathbin{\scalebox{1.5}{\ensuremath{\cap}}} \mathcal{V}(S'_0) = \emptyset \;,  \makebox[0.7in]{}{\rm if}\;\; S_0 \neq S'_0 \;,
\end{align}
where the labels $S_0$ are representatives of the classes obtained from the equivalence relation,
\begin{align}
 S' \sim S \makebox[.5in]{\rm iff}  S' = U S\;, \makebox[0.7in]{}{\rm  with~regular}\;\;U\;.  
\end{align}
In particular, the different $\mathcal{V}(S_0)$ cannot contain physically equivalent gauge fields.    Therefore, the partition function and the average of an observable $O$ over $\{ \mathcal{A}_\mu\}$ become a sum over partial contributions  
\begin{align} 
   & Z_{\rm YM}=\sum_{S_0} Z_{(S_0)}\;,&   \langle O \rangle_{\rm YM}=\sum_{S_0} \langle O \rangle_{(S_0)} \,  \frac{Z_{(S_0)}}{Z_{\rm YM}}  \;,
    \label{ZYM-part} \\&
   Z_{(S_0)} = \int_{\mathcal{V}(S_0)}[D\mathcal{A}_\mu] \, e^{-S_{\rm YM}}\;, &   \langle O \rangle_{(S_0)} = \frac{1}{ Z_{(S_0)}} \int_{\mathcal{V}(S_0)} [D\mathcal{A}_\mu] \, O\,  e^{-S_{\rm YM}}  \;. 
\end{align}
Configurations with regular $S[\mathcal{A}]$ are in the same sector, which may be labeled by  $S_0 \equiv I$. 
This sector is expected to be  the relevant one for describing weakly interacting particles in the UV regime. On the other hand, in the region around a distribution of closed surfaces, smooth variables $\mathcal{A}_\mu$ may
involve large gauge transformations with multivalued angles in their formulation.  
Accordingly, $S[\mathcal{A}]$ and the associated choice of $S_0$   is characterized by a distribution of center vortices and correlated monopoles with non-Abelian degrees of freedom (d.o.f.) \cite{oxman4d}.  
Then, from this perspective, the Singer no-go theorem appears as the foundational basis for the YM theory to give place to a YM ensemble. 
Concerning the possible relation with quark confinement,  center vortices have been considered  as relevant degrees to describe the infrared properties of YM theory 
\cite{thooft,mandelstam,nambu}.
In the lattice, at asymptotic distances, they account for a Wilson loop area law with $N$-ality
 \cite{DelDebbio:1996lih,Reinhardt:2001kf,
Engelhardt:1999xw,Langfeld:1997jx,DelDebbio:1998luz,Faber:1997rp,deForcrand:1999our,
Ambjorn:1999ym,Engelhardt:1999fd,Bertle:2001xd,Gattnar:2004gx}.   
More recently, phenomenological ensembles containing the possible defects that characterize $S_0$  led to effective models that can accommodate the properties of the confining flux tube \cite{oxman4d}.  

The aim of this work is to advance towards the establishment of a Yang-Mills ensemble from first principles. In this respect, renormalizability of each sector $Z_{(S_0)}$ is important to have well-defined calculable partial contributions. In Ref. \cite{Oxman:2015ira}, we showed that each 
 $\mathcal{V}(S_0)$ has its own BRST transformation. Although the algebraic structure does not change from sector to sector, the 
regularity conditions of the ghosts do change. This is needed in order for the regularity conditions, 
at the defects of $S_0$, to be BRST invariant. That is, the field modes to expand $\psi_I$
are inherited by the ghosts.
Here, we were able to prove the all-order multiplicative renormalizability of the center-vortex free sector, which we expect to be essentialy perturbative. In other sectors,   most of the Ward identities remain valid, but the ghost equation should be modified by sector-dependent terms. Additionaly, new counterterms arise located at the center-vortex guiding centers. The associated difficulties for establishing the renormalizability of these sectors will be dealt with in a future contribution.

%----------------------------------------------------------------------------------------
\section{The YM quantization on the $\mathcal{V}(S_0)$-sectors}
\label{sec-q}
%--------- -------------------------------------------------------------------------------

As proposed in \cite{Oxman:2015ira}, the correlation between the gauge field $\mathcal{A}_\mu$ and the phase $S(\mathcal{A})$, used to fix the gauge on each sector, can be done by means of the solutions to the equations of motion 
\begin{align}
&  \frac{\delta S_{\rm aux}}{\delta \psi_I^a} =0 \makebox[.5in]{,} S_{\rm aux} = \int_x \left( \frac{1}{2}  D_\mu^{ab} \psi_I^b D_\mu^{ac}\psi_I^c+V_{\rm aux} \right) \makebox[.5in]{,}
& D_\mu^{ab} &\equiv &\delta^{ab}\partial_\mu +gf^{acb}A_\mu^c\;. 
\label{Scalar_Action} 
\end{align}
We  initially  choose the auxiliary potential $V_{\rm aux}(\psi)$  to be the most general one constructed in terms of antisymmetric structure constants, and containg up to quartic terms. Then, besides color symmetry, we take $I= 1, \dots, N^2-1$ and impose adjoint flavor symmetry:
\begin{align} 
    V_{\rm aux} (\psi)=\frac{\mu^2}{2} \psi_I^a\psi_I^a+\frac{\kappa}{3} f^{abc}f_{IJK}\psi_I^a\psi_J^b\psi_K^c+\frac{\lambda}{4} \gamma^{abcd}_{IJKL}\psi_I^a\psi_J^b\psi_K^c\psi_L^d\;,
\end{align}
where $\gamma$ is general a color and flavor invariant tensor, \emph{i.e.}, 
\begin{eqnarray}
   &R^{aa'}R^{bb'}R^{cc'}R^{dd'}\gamma^{a'b'c'd'}_{IJKL}=\gamma^{abcd}_{IJKL}\;, \nonumber\\
    &R^{II'}R^{JJ'}R^{KK'}R^{LL'}\gamma^{abcd}_{I'J'K'L'}=\gamma^{abcd}_{IJKL}\; \;,\;
    R \in Ad(SU(N)) \label{inv}\;, 
\end{eqnarray}
composed of combinations of antisymmetric structure constants and Kronecker deltas, while $\mu$ and $\kappa$ are mass parameters. In order for the procedure to be well-defined, a potential with minima displaying $SU(N) \to Z(N)$ is essential. This pattern, which can be easily accommodated by $N^2-1$ flavors \cite{Oxman:2012ej}, produces a strong correlation between $\mathcal{A}_\mu$ and the local phase $S[{\mathcal{A}}]$. As in Ref. 
\cite{Oxman:2015ira}, a tuple $q_I = S^{-1} \psi_I S$ is called the modulus of $\psi_I$, if
$S\in SU(N)$ minimizes $( q_I  - u_I)^2 $,  
where $u_I \in \mathfrak{su}(N)$ is a reference tuple of linearly independent vectors. Then, if we perform an infinitesimal rotation of $q_I$ with generator $X$, we get $ ( q_I  - u_I, [q_I ,X]) =0$
for every $X \in \mathfrak{su}(N)$. That is, the polar decomposition and the modulus condition become 
\begin{equation}
\psi_I  = S q_I S^{-1}  \makebox[.5in]{,}  [u_I ,q_I] =0  \;.
\label{form}
\end{equation}
At the quantum level, to filter the configurations $\mathcal{A}_\mu \in \mathcal{V}(S_0)$, an identity was constructed from a Dirac delta functional over the scalar field equations of motion, together with flavored ghosts $c_I$ \cite{Oxman:2015ira}. Then, $\psi_I$ was restricted to have a polar decomposition with a local phase of the form $S= U S_0$, where $U$ is regular and $S_0$ contains center-vortex and correlated monopole defects. In particular, by using a reference tuple $u_I$ in the vacua manifold, we require the asymptotic condition $q_I \to u_I$, plus appropriate regularity conditions at the defects of $S_0$ (see Sec. 3.1). Next, the gauge-fixing in this sector was implemented by changing variables to $A_\mu$,
 $\mathcal{A}_\mu = A_\mu^U$. Thus, in the gauge-fixed expressions we can write $A_\mu$ in the place of $\mathcal{A}_\mu$, and 
  \begin{gather} 
  \zeta_I\equiv S_0 q_IS_0^{-1}\;,
  \label{zet}
  \end{gather}
in the place of $\psi_I$. The presence of $S_0$ occurs because, when performing a shift of group variables within the Fadeev-Popov procedure, it is impossible to eliminate $S_0$ with a regular gauge transformation. The pure modulus condition can also be written as 
$[\eta_I , \zeta_I] =0$, using the classical (background) field $\eta_I\equiv S_0 u_I S_0^{-1}$. 

The full Yang Mills action in this gauge is then given by
\begin{eqnarray}
\Sigma &=& S_{\rm YM}+ \int_x \left\{
\left( D_\mu^{ab}\bar{c}_I^b\right) D_\mu^{ac} c_I^c + \left( D_\mu^{ab}b_I^b\right) D_\mu^{ac} \zeta_I^c
 +\kappa f_{IJK}f^{abc}\Big( b_I^a \zeta_J^b \zeta^c_K -2\bar{c}_I^a \zeta^b_K c^c_J \Big)+
\right.\nonumber\\ 
&+& 
 \lambda\gamma_{IJKL}^{abcd}(b_I^a\zeta_J^b\zeta_K^c\zeta_L^d+3\bar{c}_I^ac_J^b\zeta_K^c\zeta_L^d)
 + \mu^2 \Big(\bar{c}^a_I c^a_I + b^a_I \zeta^b_I\Big)+
 \nonumber\\
 &+&\left. if^{abc}b^a\eta_I^b\zeta_I^c + f^{ecd}f^{eba}\bar{c}^a\eta_I^b\zeta_I^cc^d
 +if^{abc}\bar{c}^a\eta_I^bc_I^c \right\}\;,\label{Action_0}
\end{eqnarray}
\begin{eqnarray}
S_{\rm YM} &=& \frac{1}{2}\int_x\left[ 
(\partial_\mu A_\nu^a) ^2 -\partial_\mu A_\nu^a\partial_\nu A_\mu^a 
+g f^{abc}A_\mu^aA_\nu^b(\partial_\mu A_\nu^c - \partial_\nu A_\mu^c)
+\frac{g^2}{2} f^{abc}f^{dec}A_\mu^a A_\nu^b A_\mu^d A_\nu^e\right]\;.\nonumber\\
\end{eqnarray}
The action \eqref{Action_0} is invariant under the following BRST transformations \cite{Oxman:2015ira}
\begin{eqnarray} 
sA_\mu^a &=& \frac{i}{g}D^{ab}_\mu c^b \;,\;\;\;\;\;\;\;\;\;\;\;\;\;\; sc\;\;\;=\;\;\;-\frac{i}{2}f^{abc}c^b c^c\;,\nonumber\\
s\bar{c}^a&=&-b^a \;,\;\;\;\;\;\;\;\;\;\;\;\;\;\;\;\;\;\;sb^a\;\;\;=\;\;\;0\;,\nonumber\\
s\zeta_I^a&=&if^{abc}\zeta_I^b c^c+c_I^a \;,\;\;\;s\bar{c}^a_I\;\;\;=\;\;\;-if^{abc}\bar{c}^b_I c^c- b^a_I\;,\nonumber\\
sb_I&=&if^{abc}b^b_I c^c  \;,\;\;\;\;\;\;\;\;\;\;\;  sc^a_I \;\;\;=\;\;\; -if^{abc}c^b_I c^c \,.
\label{BRST_0}
\end{eqnarray}
 Conveniently, the gauge fixing terms can be written as a BRST-exact term, so that  the action \eqref{Action_0} is equivalent to
\begin{equation}
 S=S_{\rm YM} - s\int_x \Big[ D_\mu^{ab} \bar{c}_I^bD_\mu^{ac} \zeta_I^c
 + \bar{c}_I^a\Big( \mu^2 \zeta_I + \kappa f^{IJK} f^{abc}\zeta_J^b \zeta_K^c  \Big)+\gamma^{abcd}_{IJKL}\lambda \bar{c}_I^a\zeta_J^b\zeta_K^c\zeta_L^d+if^{ abc}\bar{c}^a\eta_I^b\zeta_I^c
 \Big]\;.
 \end{equation}

%-----------------------------------
\subsection{Some remarks about the BRST invariance}
%-----------------------------------

Let us focus on  the flavor-sector, which implements the correlation between the gauge field and the auxiliary adjoint scalar fields
\begin{eqnarray}
\tilde{S}_{\rm f} &= &S_{\rm YM} +\int_x \Big[
  D_\mu^{ab}\bar{c}_I^bD_\mu^{ac} c_I^c + \mu^2 \Big(\bar{c}_I^ac_I^a +b_I^a \zeta_I^a\Big)
 +\kappa f_{IJK}f^{abc}\Big(b_I^a\zeta_J^b \zeta_K^c-2 \bar{c}_I^a\zeta_K^b c_J^c \Big)+
\nonumber \\ 
&& 
+ \lambda\gamma_{IJKL}^{abcd}(b_I^a\zeta_J^b\zeta_K^c\zeta_L^d+3\bar{c}_I^ac_J^b\zeta_K^c\zeta_L^d)
  +D_\mu^{ab} b_I^b D_\mu^{ac} \zeta_I^c\Big]\;,
 \label{Sf_1}
\end{eqnarray}
which can be written as
\begin{eqnarray}
\tilde{S}_{\rm f}  &= &S_{\rm YM} - s\,\int_x  \bar{c}_I^a\frac{\delta S_H}{\delta \psi_I^a}\Big |_{\psi =\zeta_I}
\nonumber\\
&=&S_{\rm YM} - s\,\int_x \left[D_\mu^{ab}\bar{c}_I^bD_\mu^{ac} \zeta_I^c
+\bar{c}^a\frac{\delta V_H}{\delta \psi_I^a}\Big |_{\psi =\zeta_I} 
\right]\;,
\label{Sf_2}
\end{eqnarray}
with
\begin{equation}
\frac{\delta V_H}{\delta \psi_I^a}\Big |_{\psi =\zeta_I} = \mu^2 \zeta_I^a + \kappa f^{IJK}f^{abc}\zeta_J^b\zeta_K^c + \lambda \gamma_{IJKL}^{abcd}\zeta_J^b\zeta_K^c\zeta_L^d\;.
\end{equation}
As discussed in \cite{Piguet:1984js,Piguet:1995er}, in order to control the gauge-parameter independence of invariant correlation functions, it is convenient to extend the action of the BRST operator $s$ on the
gauge-fixinig parameters as
\begin{eqnarray}
s\mu^2 &=& U^2\;,\qquad sU^2\;\;\;=\;\;\;0\;, \nonumber\\
s\kappa &=& \mathcal{K}\;,\qquad \;\;\; s\mathcal{K}\;\;\;=\;\;\;0\;, \nonumber\\
s\lambda &=& \Lambda\;, \qquad \;\;\;\; s\Lambda \;\;\;=\;\;\;0\;,\label{DoubletParameters}
\end{eqnarray}
where $(U^2,\mathcal{K},\Lambda)$ are constant Grassmann parameters with ghost number 1 and mass dimension 2, 1 and zero, respectively. Thus, since physical quantities must belong to the (nontrivial) cohomology of the $s$-operator, all the gauge-fixing parameters will be on the trivial sector of the cohomology of this operator, in accordance with the BRST doublet theorem \cite{Piguet:1995er}. The complete classical action $S_{\rm f} $ is given by
\begin{eqnarray}
S_{\rm f}  = \tilde{S}_{\rm f}  - \int_x\Big(U^2 \bar{c}_I^a \zeta_I^a
+\mathcal{K}f^{IJK} f^{abc} \bar{c}_I^a\zeta_J^b \zeta_K^c
+\Lambda\gamma_{IJKL}^{abcd}\bar{c}_I^a\zeta_J^b\zeta_K^c\zeta_L^d \Big)\;,
\label{Sf_2}
\end{eqnarray}
which is invariant under the extended transformations \eqref{DoubletParameters}.
The full action in the flavor-sector is the following 
\begin{eqnarray}
\Sigma_{\rm I-sec} &=& S_{\rm YM}+  \int_x\Big\{
\left( D_\mu^{ab}\bar{c}_I^b\right) D_\mu^{ac} c_I^c + \left( D_\mu^{ab}b_I^b\right) D_\mu^{ac} \zeta_I^c
+ \mu^2 \Big(\bar{c}^a_I c^a_I + b^a_I \zeta^b_I\Big)+
\nonumber\\ 
&+&
 \kappa f_{IJK}f^{abc}\Big( b_I^a \zeta_J^b \zeta^c_K -2\bar{c}_I^a \zeta^b_K c^c_J \Big)
+ \lambda\gamma_{IJKL}^{abcd}(b_I^a\zeta_J^b\zeta_K^c\zeta_L^d+3\bar{c}_I^ac_J^b\zeta_K^c\zeta_L^d)+
\nonumber \\ 
&-&U^2 \bar{c}^a_I \zeta^a_I 
-\mathcal{K}f^{IJK} f^{abc} \bar{c}^a_I \zeta^b_J \zeta^c_K
-\Lambda \gamma^{abcd}_{IJKL}\bar{c}^a_I \zeta^b_J \zeta^d_I \zeta^e_J \Big\}\;.
\end{eqnarray}

%%%%%%%%%%%%%%%%%%%%%%%%%%%%%%%%%%%%%%%%%%
\section{The complete action and its symmetries}
%%%%%%%%%%%%%%%%%%%%%%%%%%%%%%%%%%%%%%%%%%

%\subsection{Extended action}

Owing to the nonlinearity of the BRST transformations \eqref{BRST_0}, the renormalization of some composite operators is necessary. With this purpose, we need to couple a source to each nonlinear variation of the fields, which can be done in a BRST-exact manner by imposing the $s$-variation of all these sources to be zero, \emph{ i.e.},
\begin{eqnarray}
\Sigma_{\mathrm{sources}}^{(1)}& = & \int_x\left[K^{a}_\mu (sA_\mu^a)+\bar{C}^a (sc^a) + Q_I^a (s\zeta_I^a)
+\bar{L}_I^a(s c_I^a) +  L_I^a (s\bar{c}_I^a)+ B_I^a (sb_I^a)\right]
\nonumber\\
&=& \int_x\left[\frac{i}{g} K^{a}_\mu D^{ab}_{\mu}c^b-\frac{1}{2}i\bar{C}^a f^{abc}c^{b}c^{c} 
+Q_I^a (if^{abc}\zeta_I^bc^c+c_I^a) -if^{abc}\bar{L}_I^ac_I^bc^c +
\right.\nonumber\\
&-& \left.   L_I^a (if^{abc}\bar{c}_I^bc^c+b_I^a)
+if^{abc} B_I^a b_I^bc^c\right]\;,
\end{eqnarray}
with $s(K^{a}_\mu,\bar{C}^a,Q_I^a,\bar{L}_I^a,L_I^a,B_I^a)=0$. An additional pair of external sources $(M^{ab}_I,N^{ab}_I)$, to form a convenient composite operator, will also be necessary
\begin{eqnarray}
\Sigma_{\mathrm{sources}}^{(2)}& = & s\int_x M^{ab}_I \bar{c}^b \zeta^b_I= \int_x \left(N^{ab}_I \bar{c}^a \zeta^b_I - M^{ab}_I b^a \zeta^b_I - M^{ab}_I\bar{c}^a\frac{\delta\Sigma}{\delta Q_I^b} \right)\;.
\label{ExtSources2}
\end{eqnarray}
These sources will be used to restore the Ward identities associated to the ghost and anti-ghost equations, a key step towards the algebraic proof of renormalizability. After the introduction of the external sources, for the complete extended classical action $\Sigma$ one has 
\begin{eqnarray}
\Sigma &=& S_{\rm YM}+\int_x\left[\left( D_\mu^{ab}\bar{c}_I^b\right) D_\mu^{ac} c_I^c + \left( D_\mu^{ab}b_I^b\right) D_\mu^{ac} \zeta_I^c
 +\kappa f_{IJK}f^{abc}\Big( b_I^a \zeta_J^b \zeta^c_K -2\bar{c}_I^a \zeta^b_K c^c_J \Big)+
\right.\nonumber\\ 
&+& \mu^2 \Big(\bar{c}^a_I c^a_I + b^a_I \zeta^b_I\Big)
+ \lambda\gamma_{IJKL}^{abcd}(b_I^a\zeta_J^b\zeta_K^c\zeta_L^d+3\bar{c}_I^ac_J^b\zeta_K^c\zeta_L^d)+ \nonumber\\
 &-&
 U^2 \bar{c}^a_I \zeta^a_I -\mathcal{K}f^{IJK} f^{abc} \bar{c}^a_I \zeta^b_J \zeta^c_K
-\Lambda f^{abc}f^{cde}\bar{c}^a_I \zeta^b_J \zeta^d_I \zeta^e_J 
 +if^{abc} \left(b^a\eta_I^b\zeta_I^c +\bar{c}^a\eta_I^bc_I^c \right) +
 \nonumber\\
&+& f^{ecd}f^{eba}\bar{c}^a\eta_I^b\zeta_I^cc^d
+\frac{i}{g} K^a_\mu(D_\mu^{ab}c^b)
-\frac{1}{2}i\bar{C}^a f^{abc}c^{b}c^{c} 
+Q_I^a (if^{abc}\zeta_I^bc^c+c_I^a) -if^{abc}\bar{L}_I^ac_I^bc^c +
\nonumber\\
&-&  L_I^a (if^{abc}\bar{c}_I^bc^c+b_I^a)
+if^{abc} B_I^a b_I^bc^c
+N^{ab}_I \bar{c}^a \zeta^b_I - M^{ab}_I b^a \zeta^b_I - M^{ab}_I\bar{c}^a\frac{\delta\Sigma}{\delta Q_I^b}  \Bigg]\;,
\label{FullAction}
\end{eqnarray}
which is invariant under the full set of BRST transformations 
\begin{eqnarray}
sA_\mu^a &=& \frac{i}{g}D^{ab}_\mu c^b \;,\;\;\;\;\;\;\;\;\;\;\;\;\;\; sc\;\;\;=\;\;\;-\frac{i}{2}f^{abc}c^b c^c\;,\nonumber\\
s\bar{c}^a&=&-b^a \;,\;\;\;\;\;\;\;\;\;\;\;\;\;\;\;\;\;\;sb^a\;\;\;=\;\;\;0\;,\nonumber\\
s\zeta_I^a&=&if^{abc}\zeta_I^b c^c+c_I^a \;,\;\;\;s\bar{c}^a_I\;\;\;=\;\;\;-if^{abc}\bar{c}^b_I c^c- b^a_I\;,\nonumber\\
sb_I&=&if^{abc}b^b_I c^c  \;,\;\;\;\;\;\;\;\;\;\;\;  sc^a_I \;\;\;=\;\;\; -if^{abc}c^b_I c^c\;,\nonumber \\
s\mu^2 &=& U^2 \;,\;\;\;\;\;\;\;\;\;\;\;\;\;\;\;\;\;\;\; sU^2\;\;\;=\;\;\;0\;,\nonumber\\
s\kappa &=& \mathcal{K}\;,\;\;\;\;\;\;\;\;\;\;\;\;\;\;\;\;\;\;\;\;\;\; s\mathcal{K}\;\;\;=\;\;\;0\;,\nonumber\\
s\lambda &=& \Lambda \;,\;\;\;\;\;\;\;\;\;\;\;\;\;\;\;\;\;\;\;\;\;\; s\Lambda \;\;\;=\;\;\;0\;,\nonumber\\
sM^{ab}_I&=&N_I^{ab} \;,\;\;\;\;\;\;\;\;\;\;\;\;\;\;\; sN_I^{ab} \;\;\;=\;\;\; 0\;,\nonumber\\
s\bar{C}^a&=&sK_\mu^a\;\;\;=\;\;\;sL_I^a\;\;\;=\;\;\;s\bar{L}^a_I\;\;\;=\;\;\;sQ^a_I\;\;\;=\;\;\;sB^a_I\;\;\;=\;\;\;0\;.
\label{fullBRST}
\end{eqnarray}

%\subsection{Ward identities}
\subsection{Regularity conditions in sectors labeled by magnetic defects}
 
As an example, let us consider a sector labeled by a center-vortex, In this case, the singular gauge transformation $S_0$ may be given by $S_0=e^{i\chi 2N\omega_p T_p}$, where $\omega$ being a weight of the fundamental representation (see \cite{Oxman:2015ira} and refs. therein). The generators $T_p$, $p=1, \dots, N-1$, belong to  the Cartan sector of $\mathfrak{su}(N)$, while $\chi$ is multivalued when we go around the vortex worldsheet.  The  color components of the field $q_I$ are defined by $q_I=q_I^aT_a$, $a=1\dots N^2-1$. The Lie basis consists of $N-1$ Cartan elements $T_q$, and a pair of elements $E_\alpha$, $E_{-\alpha}=E_\alpha^\dagger$ for each positive root $\alpha$ of $\mathfrak{su}(n)$. To compute $\zeta_I$ in Eq. \eqref{zet}, we can use 
 \begin{align}
    & S_0 T_q S_0^{-1}=T_q\;, \\&
     S_0 E_\alpha S_0^{-1}= \cos(2 N\omega\cdot\alpha)\chi E_\alpha+ \sin(2 N\omega\cdot\alpha)\chi E_\alpha \;.
 \end{align}
Therefore, to ensure regularity, the components $q_I^\alpha$ such that $\alpha\cdot\beta \neq 0$ must vanish on the vortex worldsheet. However, as argued in \cite{Oxman:2015ira},  we must make sure that this regularity condition is invariant under BRST transformations. That is, for these roots we must impose $s q_I^\alpha=0$ on the vortex worldsheet. A way to impose these conditions is to add 
 to the action the following term
\begin{align}
    S_{\rm b.c.}=\int_x \int d\sigma_1d\sigma_2\; \delta(x-\bar{x}(\sigma_1,\sigma_2))(\lambda_I^\alpha \zeta_I^{\alpha}+\xi_I^\alpha s\zeta_I^{\alpha})= \int_x J(\lambda_I^\alpha \zeta_I^{\alpha}+\xi_I^\alpha s\zeta_I^{\alpha})\;.
\end{align}
The fields $\xi_I^\alpha$,$\lambda_I^\alpha$ are Lagrange multipliers satisfying $s\lambda_I^\alpha=0$, $s\xi_I^\alpha=-\lambda_I^\alpha$, and $\bar{x}(\sigma_1,\sigma_2)$ is a parametrization of the vortex worldsheet. To account for a general sector, we should consider a general source $J(x)$ localized on te various magnetic defects. Using these BRST transformations, we can write $S_{\rm b.c.}=\int_x s(J \xi_I^\alpha \zeta_I^{\alpha})$, and the full action in this sector turns out to be 
\begin{align}
    \Sigma^{S_0}=\Sigma+S_{\rm b.c.}=s(\rm{something})\;, \label{actions0}
\end{align}
 where $\Sigma$ was defined in Eq. \eqref{FullAction}. 
 This is an important construction, since the Quantum Action Principle \cite{Lowenstein:1971vf,Lowenstein:1971jk,Lam:1972mb,Lam:1973qa,Clark:1976ym} can be applied to the action in this form.

\subsection{Ward identities}

In the center-vortex free sector, the action and BRST transformations are those of eqs. \eqref{FullAction}, \eqref{fullBRST}, respectively, with $\zeta_I$ and $\eta_I$ replaced by $q_I$ and $u_I$, respectively. We now display the rich set of Ward identities enjoyed by this action:
\begin{itemize}
\item{The Slavnov-Taylor identity:
\begin{eqnarray}
\nonumber S(\Sigma) &=& \int_x \left(
\frac{\delta\Sigma}{\delta K^a_\mu}\frac{\delta\Sigma}{A^a_\mu} 
+  \frac{\delta\Sigma}{\delta\bar{L}^a_I}\frac{\delta\Sigma}{\delta c_I^a}  
+ \frac{\delta\Sigma}{\delta L^a_I} \frac{\delta\Sigma}{\delta \bar{c}_I^a} 
 +\frac{\delta\Sigma}{\delta Q^a_I}\frac{\delta\Sigma}{\delta q_I^a} 
+\frac{\delta\Sigma}{\delta B^a_I}\frac{\delta\Sigma}{\delta b_I} 
 + \frac{\delta\Sigma}{\delta\bar{C}^a}\frac{\delta\Sigma}{\delta c^a}+
\right.\nonumber\\
&-&\left.
   b^a\frac{\delta\Sigma}{\delta\bar{c}^a}
  +N_I^{ab}\frac{\delta\Sigma}{\delta M_I^{ab}}
 \right)
 +U^2\frac{\delta \Sigma}{\delta\mu^2} 
 +\mathcal{K}\frac{\delta \Sigma}{\delta\kappa} 
 +\Lambda\frac{\delta \Sigma}{\delta\lambda}=0\;.
\label{ST}
\end{eqnarray}
In view of the algebraic characterization of the counterterm, we introduce the so-called linearized Slavnov-Taylor operator $\mathcal{B}_{\Sigma}$ \cite{Piguet:1995er} defined as
\begin{eqnarray}
\mathcal{B}_{\Sigma} &=& \int_x \left(
\frac{\delta\Sigma}{\delta K^a_\mu}\frac{\delta}{A^a_\mu} +  \frac{\delta\Sigma}{\delta A^a_\mu}\frac{\delta}{\delta K^a_\mu} +  \frac{\delta\Sigma}{\delta\bar{L}^a_I}\frac{\delta}{\delta c_I^a} +  \frac{\delta\Sigma}{\delta c^a_I}\frac{\delta}{\delta \bar{L}^a_I} 
 + \frac{\delta\Sigma}{\delta L^a_I} \frac{\delta}{\delta \bar{c}_I^a} 
 +  \frac{\delta\Sigma}{\delta \bar{c}^a_I}\frac{\delta}{\delta L^a_I} +
\right.\nonumber\\
&+& 
 \frac{\delta\Sigma}{\delta Q^a_I}\frac{\delta}{\delta q_I^a} 
 +  \frac{\delta\Sigma}{\delta q^a_I}\frac{\delta}{\delta Q^a_I}
+\frac{\delta\Sigma}{\delta B^a_I}\frac{\delta}{\delta b_I} +  \frac{\delta\Sigma}{\delta b^a_I}\frac{\delta}{\delta B^a_I}  + \frac{\delta\Sigma}{\delta\bar{C}^a}\frac{\delta}{\delta c^a}  
+ \frac{\delta\Sigma}{\delta c^a}  \frac{\delta}{\delta\bar{C}^a}+
\nonumber\\
&-& \left.
b^a\frac{\delta}{\delta\bar{c}^a}
 +N_I^{ab}\frac{\delta}{\delta M_I^{ab}}
\right)
 +U^2\frac{\delta }{\delta\mu^2} 
 +\mathcal{K}\frac{\delta }{\delta\kappa} 
 +\Lambda\frac{\delta }{\delta\lambda} \;,
\label{STOp}
\end{eqnarray}}
which is nilpotent, $\mathcal{B}_{\Sigma}^2=0$.

\item {Gauge fixing condition
\begin{equation}
\frac{\delta\Sigma}{\delta b^a}=if^{abc}u^b_Iq^c_I  -M^{ab}_Iq_I^b\;.
\label{WI_gfEq}
\end{equation}}
 %Note that the equation of motion of the Nakanishi-Lautrup field $b$, namely the gauge-fixing equation, does  correspond to a Ward identity of the model, because of the \textit{apparent} non-linearity of the gauge condition, \textit{i.e.}
%\begin{equation}
%\frac{\delta\Sigma}{\delta b^a}=if^{abc}u^b_Iq^c_I 
%\end{equation}
%is in fact a linearly broken equation, since $u_I$ is a fixed background-like field in terms of which the Faddeev-Popov condition is implemented.
\item{The antighost equation
\begin{equation}
\bar{\mathcal{G}}^a\Sigma
=N_{I}^{ab}q_I^b\;,
\end{equation}
with the anti-ghost operator given by
\begin{equation}
\bar{\mathcal{G}}^a=\frac{\delta}{\delta\bar{c}^a}+M_I^{ab}\frac{\delta}{\delta Q_I^b}-if^{abc}u_I^b\frac{\delta}{\delta Q_I^c} \;.
\end{equation}}
\item{The ghost equation
\begin{eqnarray}
\mathcal{G}^a\Sigma&=&
if^{abc}\left( \bar{C}^{b}c^c + Q^{b}_I q_I^c + \bar{L}^{b}_I c^c_I + L^{b}_I \bar{c}^c_I+ B^{b}_I b^c_I \right) 
+\frac{i}{g}D_\mu^{ab}K_\mu^{b}\;,
\label{WI_ghostEq}
\end{eqnarray}
with the ghost operator given by
\begin{eqnarray}
\mathcal{G}^a&=& \frac{\delta}{\delta c^a} - f^{abc}\,f^{cmn}u_I^n\frac{\delta}{\delta N_I^{mb}} \;.
\end{eqnarray}} 
\item{Ghost number equation:
\begin{eqnarray}
\mathcal{N}_{gh}\Sigma
&=&0\;,
\end{eqnarray}
\begin{eqnarray}
\mathcal{N}_{gh} &=&\int d^{4}x\,\bigg(
c^{a}_I\frac{\delta}{\delta c^{a}_I} -\bar{c}^{a}_I\frac{\delta}{\delta \bar{c}^{a}_I}
+c^{a}\frac{\delta}{\delta c^{a}} -\bar{c}^{a}\frac{\delta}{\delta \bar{c}^{a}}
+U^{2}\frac{\delta}{\delta U^{2}}
+\mathcal{K}\frac{\delta}{\delta \mathcal{K}}
+\Lambda\frac{\delta}{\delta \Lambda}+
\nonumber\\
&-&
K^{a}\frac{\delta}{\delta K^a}-2\bar{C}^{a}\frac{\delta}{\delta \bar{C}^a}
-2\bar{L}^{a}_I\frac{\delta}{\delta \bar{L}^{a}_I}
-Q^{a}_I\frac{\delta}{\delta Q^{a}_I}-B^{a}_I\frac{\delta}{\delta B^{a}_I}
 +N_I^{ab}\frac{\delta}{\delta N_I^{ab}}
\bigg)\;.
\end{eqnarray}
}
\item{Global flavor symmetry:
\begin{equation}
\mathcal{Q}\Sigma = 0 \;,
\end{equation}
where we have defined the flavor charge operator
\begin{eqnarray}
\mathcal{Q}
&\equiv &q_I^a\frac{\delta}{\delta q_I^a}-b_I^a\frac{\delta }{\delta b_I^a} 
-\bar{c}_I^a\frac{\delta }{\delta  \bar{c}_I^a}
+c_I^a\frac{\delta }{\delta c_I^a}
-u^a_I\frac{\delta }{\delta u^a_I}
-Q_I^a\frac{\delta}{\delta Q_I^a}
+B_I^a\frac{\delta}{\delta B_I^a}
+L^a_I\frac{\delta}{\delta L^a_I}+
 \nonumber\\
 &-&
 \bar{L}^a_I\frac{\delta}{\delta \bar{L}^a_I}
-\kappa\frac{\delta }{\delta \kappa}  
 - 2\lambda\frac{\delta }{\delta \lambda}
 - \mathcal{K}\frac{\delta }{\delta \mathcal{K}}
 - 2\Lambda\frac{\delta }{\delta \Lambda}
  - N_I^{ab}\frac{\delta}{\delta N_I^{ab}} - M_I^{ab}\frac{\delta}{\delta M_I^{ab}}\;.
 \end{eqnarray}
This symmetry can be used to define new   conserved quantum number in the auxiliary flavor sector, the $\mathcal{Q}$-charge. Thus, this symmetry forbids combinations of composite fields with nonvanishing $\mathcal{Q}$-charge. The corresponding values of this charge, for each field, source and parameter, is assigned in a similar way to the ghost numbers \cite{Piguet:1995er}.}

\item{Exact rigid symmetry:
\begin{equation}
\mathcal{R} \Sigma = \bar{L}_I^ac_I^a +L_I^a\bar{c}_I^a  - q_I^aQ_I^a\;,
\label{RigidSymWI}
\end{equation}
where
\begin{equation}
\mathcal{R} = \bar{c}_I^a\frac{\delta }{\delta b_I^a} 
+ q_I^a\frac{\delta }{\delta c_I^a}-if^{abc}u_I^{a}\frac{\delta}{\delta N_I^{bc}}
-B_I^a\frac{\delta }{\delta \bar{L}_I^a}  + \bar{L}_I^a\frac{\delta }{\delta Q_I^a} 
-\kappa\frac{\delta }{\delta \mathcal{K}}-2\lambda\frac{\delta }{\delta\Lambda}-M_I^{ab}\frac{\delta}{\delta N_I^{ab}}\;.
\end{equation}
}
\end{itemize}

Notice that the right-hand sides of the broken Ward identities, that is, the gauge fixing condition, the anti-ghost equation, and the ghost equation, are linear in the quantum fields. This is compatible with the Quantum Action Principle, \emph{i.e.}, they remain classical in the perturbative expansion \cite{Piguet:1995er}.
For further use, the quantum numbers of all fields, sources and parameters are displayed in the Tables \ref{table1}, \ref{table2} and \ref{table3} (B stands for bosonic and F for fermionic statistics). 

\begin{table}[H]
\centering
\begin{tabular}{|c|c|c|c|c|c|c|c|c|c|c|c|}
\hline
Fields &$A$&$b_I$&$c_I$&$\bar{c}_I$&$q_I$&$u_I$&$\bar{c}$&$c$&$b$\\
\hline\hline
Dimension &1&1&1&1&1&1&2&0&2\\
\hline
Ghost number&0&0&1&$-1$&0&0&$-1$&1&0\\
\hline
$\mathcal{Q}$-charge &0&$-1$&1&$-1$&1&$-1$&$0$&$0$&$0$\\
\hline
Nature &B&B&F&F&B&B&F&F&B\\
\hline
\end{tabular}
\caption{The field quantum numbers.}
\label{table1}
\end{table}

\begin{table}[H]
\centering
\begin{tabular}{|c|c|c|c|c|c|c|c|c|c|c|c|}
\hline
Sources &$\bar{C}$&$K_\mu$&$\bar{L}_I$&$L_I$&$Q_I$&$B_I$&$N_I$&$M_I$\\
\hline\hline
Dimension &4&3&3&3&3&3&1&1\\
\hline
Ghost number &$-2$&$-1$&$-2$&0&$-1$&$-1$&1&0\\
\hline
$\mathcal{Q}$-charge &0&0&$-1$&$1$&$-1$&$1$&$-1$&$-1$\\
\hline
Nature &B&F&B&B&F&F&F&B\\
\hline
\end{tabular}
\caption{The quantum numbers of external sources.    }
\label{table2}
\end{table}

\begin{table}[H]
\centering
\begin{tabular}{|c|c|c|c|c|c|c|}
\hline
Parameters &$\mu^2$&$\kappa$&$\lambda$&$U^2$&$\mathcal{K}$&$\Lambda$\\
\hline\hline
Dimension &2&1&0&2&1&0\\
\hline
Ghost number &0&0&0&1&1&1\\
\hline
$\mathcal{Q}$-charge &0&$-1$&$-2$&$0$&$-1$&$-2$\\
\hline
Nature &B&B&B&F&F&F\\
\hline
\end{tabular}
\caption{The quantum numbers of parameters.  }
\label{table3}
\end{table}

In a general sector, the action \eqref{actions0} satisfies most of these Ward identities, up to minor modifications. More precisely, the following WI are satisfied:
\begin{itemize}
\item{The Slavnov-Taylor identity:
\begin{eqnarray}
\nonumber S^{S_0}(\Sigma^{S_0}) &=& \int_x \left(
\frac{\delta\Sigma^{S_0}}{\delta K^a_\mu}\frac{\delta\Sigma^{S_0}}{A^a_\mu} 
+  \frac{\delta\Sigma^{S_0}}{\delta\bar{L}^a_I}\frac{\delta\Sigma^{S_0}}{\delta c_I^a}  
+ \frac{\delta\Sigma^{S_0}}{\delta L^a_I} \frac{\delta\Sigma^{S_0}}{\delta \bar{c}_I^a} 
 +\frac{\delta\Sigma^{S_0}}{\delta Q^a_I}\frac{\delta\Sigma^{S_0}}{\delta \zeta_I^a} 
+\frac{\delta\Sigma^{S_0}}{\delta B^a_I}\frac{\delta\Sigma^{S_0}}{\delta b_I} 
 \right.\nonumber\\&+&\left. \frac{\delta\Sigma^{S_0}}{\delta\bar{C}^a}\frac{\delta \Sigma^{S_0}}{\delta c^a}-
   b^a\frac{\delta\Sigma^{S_0}}{\delta\bar{c}^a}
  +N_I^{ab}\frac{\delta\Sigma^{S_0}}{\delta M_I^{ab}}
 \right)
 +U^2\frac{\delta \Sigma^{S_0}}{\delta\mu^2} 
 +\mathcal{K}\frac{\delta \Sigma^{S_0}}{\delta\kappa} 
 +\Lambda\frac{\delta \Sigma^{S_0}}{\delta\lambda}\nonumber\\&-&\lambda_I^a\frac{\delta \Sigma^{S_0}}{\delta\xi_I^a}=0\;.
\end{eqnarray}
}
\item {Gauge fixing condition
\begin{equation}
\frac{\delta\Sigma^{S_0}}{\delta b^a}=if^{abc}u^b_I\zeta^c_I  -M^{ab}_I\zeta_I^b\;,
\label{WI_gfEq}
\end{equation}}
 %Note that the equation of motion of the Nakanishi-Lautrup field $b$, namely the gauge-fixing equation, does  correspond to a Ward identity of the model, because of the \textit{apparent} non-linearity of the gauge condition, \textit{i.e.}
%\begin{equation}
%\frac{\delta\Sigma}{\delta b^a}=if^{abc}u^b_Iq^c_I 
%\end{equation}
%is in fact a linearly broken equation, since $u_I$ is a fixed background-like field in terms of which the Faddeev-Popov condition is implemented.

\item{The antighost equation. Note that the anti-ghost operator is the same of the center-vortex free sector: 
\begin{equation}
\bar{\mathcal{G}}_{S_0}^a\Sigma
=N_{I}^{ab}\zeta_I^b \makebox[.5in]{,} \bar{\mathcal{G}}_{S_0}^a=\bar{\mathcal{G}}^a \;.
\end{equation}
}
\item{Ghost number equation:
\begin{equation}
\mathcal{N}_{gh}^{S_0}\Sigma^{S_0} =0  \makebox[.5in]{,} 
\mathcal{N}^{S_0}_{gh} \equiv\mathcal{N}_{gh}-\int d^4x\, \xi_I^a\frac{\delta}{\delta \xi_I^a}\;.
\end{equation}
}
\item{Global flavor symmetry:
\begin{equation}
\mathcal{Q}^{S_0}\Sigma = 0 \makebox[.5in]{,} 
\mathcal{Q}^{S_0}\equiv \mathcal{Q}-\xi_I^a\frac{\delta}{\delta \xi_I^a}-\lambda_I^a\frac{\delta}{\delta \lambda_I^a}\;.
\end{equation}
 }
\item{Exact rigid symmetry:
\begin{equation}
\mathcal{R}^{S_0} \Sigma^{S_0} = \bar{L}_I^ac_I^a +L_I^a\bar{c}_I^a  - \zeta_I^aQ_I^a \makebox[.5in]{,} 
\mathcal{R}^{S_0} \equiv \mathcal{R}+\xi_I^a\frac{\delta}{\delta \lambda_I^a}\;.
\end{equation}
}
\end{itemize}
%----------------------------------------------------------------------------------------
\section{Renormalizability of the center-vortex free sector}
%----------------------------------------------------------------------------------------

In order to prove that the action $\Sigma$ (cf.  Eq. \eqref{FullAction}) is multiplicatively renormalizable in the center-vortex free sector, we follow the algebraic renormalization setup \cite{Piguet:1995er}. By means of the Ward identities previously derived, we characterize the most general invariant local counterterm $\Sigma^{\text{c.t.}}$, which can be freely added to the starting action $\Sigma$. According to the Quantum Action Principle \cite{Lowenstein:1971vf,Lowenstein:1971jk,Lam:1972mb,Lam:1973qa,Clark:1976ym}, $\Sigma^{\text{c.t.}}$ is an integrated local polynomial in the fields and external sources of dimension bounded by four, and has the same quantum numbers as the starting action $\Sigma$. Further, we require that the perturbed
action $\Sigma + \epsilon\Sigma^{\text{c.t.}}$  satisfy the same Ward identities and constraints of $\Sigma$ \cite{Piguet:1995er}, to the first order\footnote{The algebraic renormalization technique \cite{Piguet:1995er} is a recursive method. Hence, to show the renormalizability of a theory at first order means that the proof is valid to all orders in perturbation theory.} in the perturbation parameter $\epsilon$. In this manner, we obtain the following set of constraints:
\begin{align}
&\mathcal{B}_{\Sigma}\Sigma^{\text{c.t.}}=0\,,\quad
\bar{\mathcal{G}}^a\Sigma^{\text{c.t.}}=0\,,\quad
\mathcal{G}^a\Sigma^{\text{c.t.}}=0\,,
\nonumber
\\
&
\mathcal{N}_{gh}\Sigma^{\text{c.t.}}=0 \;,\mathcal{Q}\Sigma^{\text{c.t.}}=0\,,\quad
\mathcal{R}\Sigma^{\text{c.t.}}=0\;.
\label{CT_WI}
\end{align}
The first one means that $\Sigma^{\text{c.t.}}$ belongs to the cohomology of the nilpotent linearized operator $\mathcal{B}_\Sigma$, in the space of integrated local polynomials in the fields, sources and parameters  bounded by dimension four. From the general results on the BRST cohomolgy of Yang-Mills theories, it follows that $\Sigma^{\text{c.t.}}$ can be decomposed as
\begin{equation}
\Sigma^{\mathrm{c.t.}}=\Delta +\mathcal{B}_{\Sigma}\Delta^{(-1)}\;,
\end{equation}
where $\Delta^{-1}$ denotes a four-dimensional integrated quantity in the fields, sources and parameters with ghost number $-1$ and vanishing flavor charge. The term $\mathcal{B}_{\Sigma}\Delta^{(-1)}$ in the equation above corresponds to the trivial solution, \textit{i.e.}, to the exact part of the cohomology of the BRST operator. On the other hand, the quantity $\Delta$ identifies the nontrivial solution, namely the cohomology of $\mathcal{B}_{\Sigma}$, meaning that $\Delta \neq \mathcal{B}_{\Sigma} \tilde{\Delta} $, for some local integrated polynomial $\tilde{\Delta}$. In order for the action to be multiplicatively renormalizable, the counterterm must be reabsorbed in the classical action by a multiplicative renormalization of the fields ($\mathcal{F}$),  sources, and parameters ($\mathcal{J}$),
\begin{eqnarray}
\Sigma_0[\mathcal{F}_0,\mathcal{J}_0]+O(\epsilon^{2})&=& \Sigma[\mathcal{F},\mathcal{J}]+\epsilon\,\Sigma^{\mathrm{c.t.}}[\mathcal{F},\mathcal{J}]\,,\nonumber\\
\mathcal{F}&=&\{A_\mu, q_I,\bar{c}_I,c_I,b_I,b,\bar{c},c\}\,,\nonumber\\
\mathcal{J}&=&\{K_\mu,L_I,\bar{L}_I,Q_I,B_I,N_I,M_I,\bar{C},g,\mu^2,\kappa,\lambda,U^2,\mathcal{K},\Lambda\}\,,\label{bare-Action}
\end{eqnarray}
where the label ``0'' indicates bare quantities. By convention we choose the renormalization factors as
\begin{eqnarray}
\mathcal{F}_0&=&Z^{1/2}_{\mathcal{F}}\,\mathcal{F}=\left(1+\frac{\epsilon}{2}\,z_{\mathcal{F}}\right)\mathcal{F}\,,\nonumber\\
\mathcal{J}_0&=&Z_{\mathcal{J}}\,\mathcal{J}=\left(1+\epsilon\,z_{\mathcal{J}}\right)\mathcal{J}\,.
\end{eqnarray} 
The coefficients $\{z_{\mathcal{F}},z_{\mathcal{J}}\}$ are  linear combinations of the free parameters in the counter-term, $\{a_0,b_i\}_{i=1,...}$ (see below). By direct inspection, and with the help of Tables \ref{table1}, \ref{table2}, and \ref{table3}, one can find that the most general counter-term with vanishing $\mathcal{Q}$-charge is of the form 
\begin{equation}
\Sigma^{\mathrm{c.t.}} = a_0S_{\mathrm{YM}} +\mathcal{B}_\Sigma \Delta^{(-1)}\;,
\end{equation}
\begin{eqnarray}
\Delta^{(-1)} &=& \int_x \Big[b_1\bar{C}^ac^a+b_2K_\mu^a A_\mu^a + b_3\bar{L}^a_Ic_I^a
+ b_4f^{abc}\bar{L}^a_I q_I^b c^c + b_5^{abc} B^{a}_I M_I^{bc} + b_6L_I^ac_I^a + b_7Q_I^a q_I^a+
\nonumber\\
&+&
b_8B_I^au_I^a +b_9B_I^ab_I^a 
+b_{10}f^{abc}(\partial_\mu A_\mu^a)\bar{c}_I^b q_I^c
+ b_{11}^{abcd}  A_\mu^a A_\mu^b\bar{c}_I^c q_I^d 
+b_{12} \bar{c}_I^a \partial^2 q_I^a    + b_{13}(\partial_\mu A_\mu^a)\bar{c}^a+
\nonumber\\
&+&
 b_{14,IJKL}^{abcd}b_I^a \bar{c}_J^b q_K^c q_L^d +b_{15} f^{abc}b_I^a \bar{c}^b q_I^c
 +b_{16,IJKL}^{abcd}c_I^a \bar{c}_J^b \bar{c}_K^c q_L^d +b_{17} f^{abc}c_I^a \bar{c}^b_I \bar{c}^c+
 \nonumber\\
&+&
b_{18,IJKL}^{abcd}\bar{c}^a_I  u_J^b q_K^c q_L^d
+b_{19}f^{IJK} f^{abc}\kappa\bar{c}^a_I  q_J^b q_K^c +b_{20,IJKL}^{abcd}\lambda\bar{c}^a_I  q_J^b q_K^cq_L^d+
\nonumber\\
&+& 
b_{21}^{abcd}\bar{c}^a_I  q_I^b \bar{c}^c c^d
+b_{22} f^{abc} q_I^au_I^b \bar{c}^c 
+b_{23}f^{abc}\bar{c}_I^aq_I^b b^c +b_{24}\mu^2 \bar{c}_I^aq_I^a+
\nonumber\\
&+& 
b_{25}f^{abc}\bar{c}^aA_\mu^bA_\mu^c+b_{26}f^{abc}\bar{c}^a\bar{c}^bc^c
+b_{27,IJKL}^{abcde}\bar{c}_I^a \bar{c}_J^b q_K^c q_L^d c^e
 + b_{28}\bar{c}^ab^a+\nonumber\\
&+&
b_{29,IJKL}^{abcde}M_I^{ab}\bar{c}_J^c q_K^d q_L^e
+b_{30}^{abcd}M_I^{ab}\bar{c}^c q_I^c
\Big]\;.
\end{eqnarray}
Terms containing the parameters of the model are forbidden in $\Delta$ due to their BRST doublet structure in eq. \eqref{DoubletParameters}.  Also, as in usual Yang-Mills theories gauged in a linear covariant way \textit{\`a la Faddeev-Popov}, the linear and quadratic terms in $A_\mu$ mixed with other fields vanish because the BRST invariance. Any other possibility results in a combination of flavored  fields with mass dimension $4$ and vanishing ghost number and $\mathcal{Q}$-charge, all of them being BRST-exact forms.  Thus, we conclude that the nontrivial cohomology of the present model is the usual cohomology corresponding to the Yang-Mills theories, namely
\begin{eqnarray}
\Delta &= & a_0S_{\mathrm{YM}}\;.
\end{eqnarray}
After a long but straightforward computation, the constraints imposed by the Ward identities \eqref{CT_WI} are the following 
\begin{align}
  &b_1=\dots= b_9=b_{13}=b_{14}=b_{16}=b_{18}=b_{22}=b_{25}=\dots=b_{30}=0\;,\nonumber\\
  &
  b_{21}^{cban}=ib_{23}(f^{mcn}f^{mba}+f^{mbn}f^{cma})\;,\nonumber\\ 
  & b_{15}=b_{23}=-b_{17}=b_{24}\;,\nonumber\\
  &b_{10}=gb_{12}\;,\nonumber\\
  & b_{11}^{abcd}=-g^2f^{ac\alpha}f^{\alpha b d}b_{12}\;,\nonumber\\
    &    b_{21}^{cban}=ib_{23}(f^{mcn}f^{mba}+f^{mbn}f^{cma})\;,\nonumber\\
    & b_{30}^{cbae}=-\delta^{ca}\delta^{be}b_7\;,\nonumber\\
    &f^{mna}b_{20,IJKL}^{mbcd}+f^{mba}b_{20,IJKL}^{nmcd}+f^{mca}b_{20,IJKL}^{nbmd}+f^{mda}b_{20,IJKL}^{nbcm}=0 \; . \label{nakanishi}
\end{align}
The most general counter-term consistent with all the Ward identities is therefore
\begin{eqnarray}
    \Sigma^{c.t.}&=&\int_x\left[\frac{a_0}{2}(\partial_\mu A^a_\nu)^2-\frac{a_0}{2}\partial_\nu A^a_\mu \partial_\mu A^a_\nu+\frac{a_0}{2}gf^{abc}A_\mu^aA_\nu^b\partial_\mu A_\nu^c+\frac{a_0}{4}g^2 f^{abc} f^{cde} A_\mu^a A_\nu^b A_\mu^d A_\nu^e+\right.\nonumber\\
    &+&\left.b_{12}(\partial_\mu \bar{c}_I^a \partial_\mu c_I^a +gf^{abc}\bar{c}_I^a\partial_\mu c_I^b A_\mu^c+gf^{abc}\partial_\mu \bar{c}_I^a A_\mu^b c_I^c +g^2 f^{abe}f^{cde}A_\mu^a \bar{c}_I^bA_\mu^c c_I^d+\right.\nonumber\\
    &+&\left.
    \partial_\mu b_I^a \partial_\mu q_I^a +gf^{abc}b_I^a\partial_\mu q_I^b A_\mu^c+gf^{abc}\partial_\mu b_I^a A_\mu^b q_I^c +g^2 f^{abe}f^{cde}A_\mu^a b_I^bA_\mu^c q_I^d)+\right.\nonumber\\
    &+&\left.
b_{19}f^{IJK}f^{abc}(\mathcal{K}\bar{c}^a_I  q_J^b q_K^c-\kappa b_I^aq_J^bq_K^c-2\kappa\bar{c}_I^ac_J^bq_K^c)+\right.\nonumber\\
&+&\left.
 b_{20,IJKL}^{abcd}(\Lambda \bar{c}^a_I  q_J^b q_K^cq_L^d-\lambda b_I^aq_J^bq_K^cq_L^d-3\lambda \bar{c}_I^ac_J^bq_K^cq_L^d)+\right.\nonumber\\
 &+&\left.
 b_{24}(U^2 \bar{c}_I^aq_I^a-\mu^2 b_I^aq_I^a-\mu^2\bar{c}_I^ac_I^a)\right] \;.
  \label{Final_CT}
\end{eqnarray}
Finally,  it remains to check if this term can be reabsorbed through a multiplicative redefinition of the fields, sources, coupling constant and parameters of the starting action, according to \eqref{bare-Action}. Indeed, there is no contribution that is not already present in the original action \eqref{FullAction}. By direct inspection, we obtain the following renormalization factors
    \begin{align}
        &z_A=a_0\;, & z_g=-\frac{a_0}{2}\;,\nonumber \\
        &z_{q_I}=0\;, & z_{c_I}=0\;,\nonumber \\
        &z_{\bar{c}_I}=2 b_{12}\;, & z_{b_I}=2 b_{12}\;,\nonumber \\
        &z_{\mathcal{K}}=-b_{12}-b_{19}\;, &z_{\kappa}=-b_{12}-b_{19}\;,\nonumber \\
        &z_\Lambda= -b_{12}-b_{20}\;, & z_\lambda= -b_{12}-b_{20}\;,\nonumber\\
        &z_{U^2}=-b_{12}-b_{25}\;, & z_{\mu^2}=-b_{12}-b_{25}\;,\nonumber\\
        &z_c=0\;, & z_{\bar{c}}=0\;, \nonumber\\
        &z_b=0\;, & z_{\bar{C}}=0\;, \nonumber\\
        & z_K=-\frac{\sigma}{2}\;,
        &z_L=-b_{12}\;, \nonumber\\ 
        &z_{\bar{L}}=0\;, & z_B=-b_{12}\;,\nonumber\\ 
        & z_Q=0\;,
        &z_N=0\;,\nonumber \\
        & z_M =0 \; .\label{z}
    \end{align}
With these relations we end the proof of the algebraic renormalizability in the center-vortex free sector of the gauge-fixing proposed in Ref. \cite{Oxman:2015ira}. Particularly, for the renormalization of the gluon field and the coupling constant we obtained the relation $Z_A=Z_g^{-1}$.  Moreover, the Faddeev-Popov ghosts and flavored pair $\{q_I,c_I\}$ do not renormalize, $Z_c=Z_{\bar{c}}=Z_{q_I}=Z_{c_I}=1$.

%----------------------------------------------------------------------------------------

\section{Conclusions}
%----------------------------------------------------------------------------------------
 
In the last decades, various approaches aimed at understanding confinement have been extensively explored.
They were based on numerous theoretical ideas, Monte Carlo simulations, phenomenological and effective models. 
The Gribov-Zwanzinger scenario is among those closer to the first principles of $SU(N)$ YM theory.
In the lattice,  $SU(N)$ (Monte Carlo) configurations have also been analyzed, leading
to the detection of percolating center vortices and monopoles as dominant degrees in the infrared. On the other side, far from the YM fundaments,  
 topological solutions to effective Yang-Mills-Higgs (YMH) models successfully reproduced asymptotic properties of the confining flux tube
 \cite{Hanany_2004,AUZZI2003187,PhysRevD.71.045010} (see also \cite{Oxman:2012ej,oxmangustavo} and references therein).
Some connections have  been established between these approaches. 
Center vortices lie on the common boundary of the fundamental modular and  Gribov regions \cite{GREENSITE2005170}.
 Recently, in Ref. \cite{oxman4d}, a phenomenological ensemble  of percolating center-vortex worldsurfaces was generated by emergent gauge fields, which are the Goldstone modes in a condensate of center-vortex loops. The inclusion of monopoles with non-Abelian d.o.f. on center vortices was effectively represented by adjoint Higgs fields. These  elements were then related to effective YMH models that can accommodate confining flux tubes with $N$-ality. 

Then, we may envisage a long road that starting from YM first principles leads to an ensemble, and then from the ensemble to the confining flux tube. Regarding the possible transition from YM theory to a YM ensemble, we believe that a controlled initial step was done in Ref. \cite{Oxman:2015ira}, where a continuum version of the lattice Laplacian-type center gauges was proposed. In this formulation, the theory is defined on infinitely many sectors that give a partition of the whole configuration space. Each sector has its own BRST transformation and invariance, being labeled by a distribution of center vortices and correlated monopoles with non-Abelian d.o.f. These are precisely the above-mentioned elements needed to make contact with effective YMH models and confining flux tubes. 

In this work, another step to settle the foundations of a YM ensemble was given. Initially, we pointed to the existence of a set of nonintegrated renormalizable Ward identities that  control the dependence on the (adjoint) flavored auxiliary fields and ghosts.  External sources were also added to restore the Faddeev-Popov ghost equation and originate a new type of quantum number. Next, we proved the renormalizability to all-orders in the center-vortex free sector, by exploring its rich set of Ward identities.   In contrast to the Landau gauge, the ghost equation is not integrated, so it is more powerfull. It implies that the couterterm cannot depend on $c^a$, and that the Faddeev-Popov ghosts do not renormalize $Z_c=Z_{\bar{c}}=1$, which is a very strong non-renormalization theorem. Another consequence is that the renormalization factors for the gluon field and the coupling constant are not independent, $Z_A=Z_g^{-1}$, since the ghost equation eliminates the conterterm $A\frac{\delta \Sigma}{\delta A}$. A similar property was obtained in the Abelian sector of the maximal Abelian gauge \cite{Capri:2008ak}. 

These advances encourage further studies about the sectors labeled by
magnetic topological degrees. Here, we showed that not only the BRST transformations but also most Ward identities maintain the same algebraic structure. The only exception is the ghost equation, which should be modified by sector-dependent terms. Additionaly, the counterterms would also contain divergences located at the center-vortex guiding centers. In a future work, we will investigate renormalizability in these sectors, which would be useful to characterize an ensemble from first principles. In this regard, quantum fluctuations around configurations of thin vortices were studied in Ref. \cite{PhysRevD.68.025001}, by considering one of the possible self-adjoint extensions to treat them. In that reference, it was also pointed out that other extensions could effectively implement the physically interesting thick center-vortex case. The generalized procedure could be applied in such a context, as the field modes in center-vortex sectors satisfy nontrivial regularity conditions on the corresponding worldsheets, which is one of the properties of the thick objects. In particular, we could approach properties such as the stiffness and tension of center vortices, thus inferring the main properties of the Yang-Mills ensemble. In this manner, we could make contact with lattice-based phenomenological proposals used to describe quark confinement in terms of percolating center vortices with positive stiffness.

%----------------------------------------------------------------------------------------
\section*{Acknowledgments}
%----------------------------------------------------------------------------------------

 The Conselho Nacional de Desenvolvimento Cient\'{\i}fico e Tecnol\'{o}gico (CNPq), the Coordena\c c\~ao de Aperfei\c coamento de Pessoal de N\'{\i}vel Superior (CAPES), and the Funda\c c\~{a}o de Amparo \`{a} Pesquisa do Estado do Rio de Janeiro (FAPERJ) are acknowledged for their financial support.

\bibliographystyle{unsrt}
\bibliography{Biblio_v7}

\begin{thebibliography}{10}

\bibitem{Faddeev:1967fc}
L.~D. Faddeev and V.~N. Popov.
\newblock {Feynman Diagrams for the Yang-Mills Field}.
\newblock {\em Phys. Lett.}, 25B:29--30, 1967.

\bibitem{Gribov:1977wm}
V.~N. Gribov.
\newblock {Quantization of Nonabelian Gauge Theories}.
\newblock {\em Nucl. Phys.}, B139:1, 1978.

\bibitem{Sobreiro:2005ec}
R.~F. Sobreiro and S.~P. Sorella.
\newblock {Introduction to the Gribov ambiguities in Euclidean Yang-Mills
  theories}.
\newblock In {\em {13th Jorge Andre Swieca Summer School on Particle and Fields
  Campos do Jordao, Brazil, January 9-22, 2005}}, 2005.

\bibitem{Singer:1978dk}
I.~M. Singer.
\newblock {Some Remarks on the Gribov Ambiguity}.
\newblock {\em Commun. Math. Phys.}, 60:7--12, 1978.

\bibitem{Zwanziger:1988jt}
D.~Zwanziger.
\newblock {Action From the Gribov Horizon}.
\newblock {\em Nucl. Phys.}, B321:591, 1989.

\bibitem{Zwanziger:1989mf}
D.~Zwanziger.
\newblock {Local and Renormalizable Action From the Gribov Horizon}.
\newblock {\em Nucl. Phys.}, B323:513--544, 1989.

\bibitem{Zwanziger:1992qr}
D.~Zwanziger.
\newblock {Renormalizability of the critical limit of lattice gauge theory by
  BRS invariance}.
\newblock {\em Nucl. Phys.}, B399:477--513, 1993.

\bibitem{Dudal:2005na}
D.~Dudal, R.~F. Sobreiro, S.~P. Sorella, and H.~Verschelde.
\newblock {The Gribov parameter and the dimension two gluon condensate in
  Euclidean Yang-Mills theories in the Landau gauge}.
\newblock {\em Phys. Rev.}, D72:014016, 2005.

\bibitem{Dudal:2008sp}
D.~Dudal, J.~A. Gracey, S.~P. Sorella, N.~Vandersickel, and H.~Verschelde.
\newblock {A Refinement of the Gribov-Zwanziger approach in the Landau gauge:
  Infrared propagators in harmony with the lattice results}.
\newblock {\em Phys. Rev.}, D78:065047, 2008.

\bibitem{Dudal:2011gd}
D.~Dudal, S.~P. Sorella, and N.~Vandersickel.
\newblock {The dynamical origin of the refinement of the Gribov-Zwanziger
  theory}.
\newblock {\em Phys. Rev.}, D84:065039, 2011.

\bibitem{VANBAAL1992259}
Pierre van Baal.
\newblock More (thoughts on) gribov copies.
\newblock {\em Nuclear Physics B}, 369(1):259 -- 275, 1992.

\bibitem{Capri:2016aqq}
M.~A.~L. Capri, D.~Dudal, D.~Fiorentini, M.~S. Guimaraes, I.~F. Justo, A.~D.
  Pereira, B.~W. Mintz, L.~F. Palhares, R.~F. Sobreiro, and S.~P. Sorella.
\newblock {Local and BRST-invariant Yang-Mills theory within the Gribov
  horizon}.
\newblock {\em Phys. Rev.}, D94(2):025035, 2016.

\bibitem{Capri:2018ijg}
M.~A.~L. Capri, D.~Dudal, M.~S. Guimaraes, A.~D. Pereira, B.~W. Mintz, L.~F.
  Palhares, and S.~P. Sorella.
\newblock {The universal character of Zwanziger's horizon function in Euclidean
  Yang–Mills theories}.
\newblock {\em Phys. Lett.}, B781:48--54, 2018.

\bibitem{Capri:2015pfa}
M.~A.~L. Capri, D.~Fiorentini, and S.~P. Sorella.
\newblock {Gribov horizon and non-perturbative BRST symmetry in the maximal
  Abelian gauge}.
\newblock {\em Phys. Lett.}, B751:262--271, 2015.

\bibitem{Dudal:2010cd}
D.~Dudal, M.~S. Guimaraes, and S.~P. Sorella.
\newblock {Glueball masses from an infrared moment problem and nonperturbative
  Landau gauge}.
\newblock {\em Phys. Rev. Lett.}, 106:062003, 2011.

\bibitem{Tissier:2010ts}
M.~Tissier and N.~Wschebor.
\newblock {Infrared propagators of Yang-Mills theory from perturbation theory}.
\newblock {\em Phys. Rev.}, D82:101701, 2010.

\bibitem{Tissier:2011ey}
M.~Tissier and N.~Wschebor.
\newblock {An Infrared Safe perturbative approach to Yang-Mills correlators}.
\newblock {\em Phys. Rev.}, D84:045018, 2011.

\bibitem{Pelaez:2014mxa}
M.~Pelaez, M.~Tissier, and N.~Wschebor.
\newblock {Two-point correlation functions of QCD in the Landau gauge}.
\newblock {\em Phys. Rev.}, D90:065031, 2014.

\bibitem{Siringo:2019lmg}
F.~Siringo.
\newblock {Yang-Mills ghost propagator in linear covariant gauges}.
\newblock {\em Phys. Rev.}, D99(9):094024, 2019.

\bibitem{Siringo:2018uho}
F.~Siringo and G.~Comitini.
\newblock {Gluon propagator in linear covariant Rxi gauges}.
\newblock {\em Phys. Rev.}, D98(3):034023, 2018.

\bibitem{10.1093/ptep/ptz093}
R.~Yanagihara and M.~Kitazawa.
\newblock {A study of stress-tensor distribution around the flux tube in the
  Abelian–Higgs model}.
\newblock {\em Progress of Theoretical and Experimental Physics}, 2019(9),
  2019.

\bibitem{cosmai}
M.~Baker, P.~Cea, V.~Chelkonov, L.~Cosmai, Cuteri F., and A.~Papa.
\newblock {Isolating the confining color field in the SU(3) flux tube}.
\newblock {\em Eur. Phys. J. C}, 79:478, 2019.

\bibitem{LUSCHER1981317}
M.~Lüscher.
\newblock Symmetry-breaking aspects of the roughening transition in gauge
  theories.
\newblock {\em Nuclear Physics B}, 180(2):317 -- 329, 1981.

\bibitem{ATHENODOROU2007132}
A.~Athenodorou, B.~Bringoltz, and M.~Teper.
\newblock The closed string spectrum of su(n) gauge theories in 2+1 dimensions.
\newblock {\em Physics Letters B}, 656(1):132 -- 140, 2007.

\bibitem{Oxman:2015ira}
L.~E. Oxman and G.~C. Santos-Rosa.
\newblock {Detecting topological sectors in continuum Yang-Mills theory and the
  fate of BRST symmetry}.
\newblock {\em Phys. Rev. D}, 92:125025, 2015.

\bibitem{oxman4d}
L.~E. Oxman.
\newblock {4d ensembles of percolating center vortices and monopole defects:
  the emergence of flux tubes with N-ality and gluon confinement}.
\newblock {\em Phys. Rev. D}, 98:036018, 2018.

\bibitem{thooft}
G.~‘t Hooft.
\newblock {On the Phase Transition Towards permanent quark confinement}.
\newblock {\em Nucl. Phys. B}, 138:1, 1978.

\bibitem{mandelstam}
S.~Mandelstam.
\newblock {II. Vortices and quark confinement in non-Abelian gauge theories}.
\newblock {\em Phys. Rept.}, 23:245, 1976.

\bibitem{nambu}
Y.~Nambu.
\newblock {Strings, monopoles, and gauge fields}.
\newblock {\em Phys. Rev. D}, 10:4262, 1974.

\bibitem{DelDebbio:1996lih}
L.~Del~Debbio, M.~Faber, J.~Greensite, and S.~Olejnik.
\newblock {Center dominance and Z(2) vortices in SU(2) lattice gauge theory}.
\newblock {\em Phys. Rev. D}, 55:2298--2306, 1997.

\bibitem{Reinhardt:2001kf}
H.~Reinhardt.
\newblock {Topology of center vortices}.
\newblock {\em Nucl. Phys. B}, 628:133--166, 2002.

\bibitem{Engelhardt:1999xw}
M.~Engelhardt and H.~Reinhardt.
\newblock {Center projection vortices in continuum Yang-Mills theory}.
\newblock {\em Nucl. Phys. B}, 567:249, 2000.

\bibitem{Langfeld:1997jx}
K.~Langfeld, H.~Reinhardt, and O.~Tennert.
\newblock {Confinement and scaling of the vortex vacuum of SU(2) lattice gauge
  theory}.
\newblock {\em Phys. Lett. B}, 419:317--321, 1998.

\bibitem{DelDebbio:1998luz}
L.~Del~Debbio, M.~Faber, J.~Giedt, J.~Greensite, and S.~Olejnik.
\newblock {Detection of center vortices in the lattice Yang-Mills vacuum}.
\newblock {\em Phys. Rev. D}, 58:094501, 1998.

\bibitem{Faber:1997rp}
M.~Faber, J.~Greensite, and S.~Olejnik.
\newblock {Casimir scaling from center vortices: Towards an understanding of
  the adjoint string tension}.
\newblock {\em Phys. Rev. D}, 57:2603--2609, 1998.

\bibitem{deForcrand:1999our}
P.~de~Forcrand and M.~D'Elia.
\newblock {On the relevance of center vortices to QCD}.
\newblock {\em Phys. Rev. Lett.}, 82:4582--4585, 1999.

\bibitem{Ambjorn:1999ym}
J.~Ambjorn, J.~Giedt, and J.~Greensite.
\newblock {Vortex structure versus monopole dominance in Abelian projected
  gauge theory}.
\newblock {\em JHEP}, 02:033, 2000.

\bibitem{Engelhardt:1999fd}
M.~Engelhardt, K.~Langfeld, H.~Reinhardt, and O.~Tennert.
\newblock {Deconfinement in SU(2) Yang-Mills theory as a center vortex
  percolation transition}.
\newblock {\em Phys. Rev. D}, 61:054504, 2000.

\bibitem{Bertle:2001xd}
R.~Bertle, M.~Engelhardt, and M.~Faber.
\newblock {Topological susceptibility of Yang-Mills center projection
  vortices}.
\newblock {\em Phys. Rev. D}, 64:074504, 2001.

\bibitem{Gattnar:2004gx}
J.~Gattnar, C.~Gattringer, K.~Langfeld, H.~Reinhardt, A.~Schafer, S.~Solbrig,
  and T.~Tok.
\newblock {Center vortices and Dirac eigenmodes in SU(2) lattice gauge theory}.
\newblock {\em Nucl. Phys. B}, 716:105--127, 2005.

\bibitem{Oxman:2012ej}
L.~E. Oxman.
\newblock {Confinement of quarks and valence gluons in SU(N) Yang-Mills-Higgs
  models}.
\newblock {\em JHEP}, 03:038, 2013.

\bibitem{Piguet:1984js}
O.~Piguet and Klaus S.
\newblock {Gauge Independence in Ordinary {Yang-Mills} Theories}.
\newblock {\em Nucl. Phys.}, B253:517--540, 1985.

\bibitem{Piguet:1995er}
O.~Piguet and S.~P. Sorella.
\newblock {\em {Algebraic renormalization: Perturbative renormalization,
  symmetries and anomalies}}, volume~28.
\newblock Springer, Berlin, 1995.

\bibitem{Lowenstein:1971vf}
J.~H. Lowenstein.
\newblock {Normal product quantization of currents in Lagrangian field theory}.
\newblock {\em Phys. Rev. D}, 4:2281--2290, 1971.

\bibitem{Lowenstein:1971jk}
J.~H. Lowenstein.
\newblock {Differential vertex operations in Lagrangian field theory}.
\newblock {\em Commun. Math. Phys.}, 24:1--21, 1971.

\bibitem{Lam:1972mb}
Y.-M.~P. Lam.
\newblock {Perturbation Lagrangian theory for scalar fields: Ward-Takahasi
  identity and current algebra}.
\newblock {\em Phys. Rev. D}, 6:2145--2161, 1972.

\bibitem{Lam:1973qa}
Y.-M.~P. Lam.
\newblock {Equivalence theorem on Bogolyubov-Parasiuk-Hepp-Zimmermann
  renormalized Lagrangian field theories}.
\newblock {\em Phys. Rev. D}, 7:2943--2949, 1973.

\bibitem{Clark:1976ym}
T.~E. Clark and J.~H. Lowenstein.
\newblock {Generalization of Zimmermann's Normal-Product Identity}.
\newblock {\em Nucl. Phys. B}, 113:109--134, 1976.

\bibitem{Hanany_2004}
A.~Hanany and D.~Tong.
\newblock Vortex strings and four-dimensional gauge dynamics.
\newblock {\em Journal of High Energy Physics}, 2004(04):066--066, 2004.

\bibitem{AUZZI2003187}
R.~Auzzi, S.~Bolognesi, J.~Evslin, K.~Konishi, and A.~Yung.
\newblock Nonabelian superconductors: vortices and confinement in n=2 sqcd.
\newblock {\em Nuclear Physics B}, 673(1):187 -- 216, 2003.

\bibitem{PhysRevD.71.045010}
A.~Gorsky, M.~Shifman, and A.~Yung.
\newblock {Non-Abelian Meissner effect in Yang-Mills theories at weak
  coupling}.
\newblock {\em Phys. Rev. D}, 71:045010, 2005.

\bibitem{oxmangustavo}
L.~E. Oxman and G.~M. Sim\~oes.
\newblock {k-Strings with exact Casimir law and Abelian-like profiles}.
\newblock {\em Phys. Rev. D}, 99:016011, 2019.

\bibitem{GREENSITE2005170}
J.~Greensite, S.~Olejník, and D.~Zwanziger.
\newblock Confinement and center vortices in coulomb gauge: analytic and
  numerical results.
\newblock {\em Nuclear Physics B - Proceedings Supplements}, 141:170 -- 176,
  2005.

\bibitem{Capri:2008ak}
M.~A.~L. Capri, V.~E.~R. Lemes, R.~F. Sobreiro, S.~P. Sorella, and R.~Thibes.
\newblock {The Gluon and ghost propagators in Euclidean Yang-Mills theory in
  the maximal Abelian gauge: Taking into account the effects of the Gribov
  copies and of the dimension two condensates}.
\newblock {\em Phys. Rev.}, D77:105023, 2008.

\bibitem{PhysRevD.68.025001}
J.~D. L\"ange, M.~Engelhardt, and H.~Reinhardt.
\newblock Energy density of vortices in the schr\"odinger picture.
\newblock {\em Phys. Rev. D}, 68:025001, 2003.

\end{thebibliography}

\end{document}